%% file: main.tex
\pdfoutput=1

\documentclass[cameraready]{ecai} 

\usepackage{latexsym}
\usepackage{amssymb}
\usepackage{amsmath}
\usepackage{amsthm}
\usepackage{booktabs}
\usepackage{enumitem}
\usepackage{graphicx}
\usepackage{color}

\usepackage[table,xcdraw]{xcolor}
\usepackage{multirow}
\usepackage{amssymb}
\usepackage{ragged2e}
\usepackage{array}       
\usepackage{threeparttable} 
\usepackage{makecell}    
\usepackage{times}
\usepackage{float}
\usepackage{pgfplots}
\usepackage{mathrsfs}
\newcolumntype{P}[1]{>{\centering\arraybackslash}p{#1}}
\usepackage{supertabular}
\usepackage{longtable}
\pgfplotsset{compat=1.18, width=7.7cm}

\usepackage{subfig}
\usepackage{overpic}   
\usepackage{enumitem} 

\usepackage{mathptmx}
\usepackage{booktabs}
\usepackage[linesnumbered,ruled,vlined]{algorithm2e}
\usepackage{algorithmic} 
\usepackage{listings}
\usepackage{tcolorbox}

\usepackage{CJKutf8}

\newcommand{\BibTeX}{B\kern-.05em{\sc i\kern-.025em b}\kern-.08em\TeX}

\begin{document}

\newcommand{\toolname}{\texttt{ASMA-Tune}\xspace}


\begin{frontmatter}



\title{ASMA-Tune: Unlocking LLMs' Assembly Code Comprehension via Structural-Semantic Instruction Tuning}


\author[A]
{\fnms{Xinyi}~\snm{Wang}\footnote{Email: 2120230741@mail.nankai.edu.cn.}}
\author[B,C]
{\fnms{Jiashui}~\snm{Wang}}
\author[D]
{\fnms{Jinbo}~\snm{Su}}
\author[D]
{\fnms{Ke}~\snm{Wang}}
\author[E]
{\fnms{Peng}~\snm{Chen}}
\author[B]
{\fnms{Yanming}~\snm{Liu}}
\author[C]
{\fnms{Long}~\snm{Liu}}
\author[A]
{\fnms{Xiang}~\snm{Li}}
\author[C]
{\fnms{Yangdong}~\snm{Wang}}
\author[C]
{\fnms{Qiyuan}~\snm{Chen}}
\author[A]
{\fnms{Rongze}~\snm{Chen}}
\author[A]
{\fnms{Chunfu}~\snm{Jia}\thanks{Corresponding Author.}}

\address[A]{Nankai University}
\address[B]{Zhejiang University}
\address[C]{Ant Group}
\address[D]{Renmin University of China}
\address[E]{University of the Chinese Academy of Sciences}






\input{sec/0_abstract}    

\end{frontmatter}
\input{sec/1_intro}
\input{sec/2_related_work}
\input{sec/3_method}
\input{sec/4_experiments}
\input{sec/5_results}

\input{sec/6_conclusion}




\begin{ack}
By using the \texttt{ack} environment to insert your (optional) 
acknowledgements, you can ensure that the text is suppressed whenever 
you use the \texttt{doubleblind} option. In the final version, 
acknowledgements may be included on the extra page intended for references.
\end{ack}



\bibliography{main}

\end{document}

%% file: sec/0_abstract.tex
\begin{abstract}
Assembly code analysis and comprehension play critical roles in applications like reverse engineering, yet they face substantial challenges due to low information density and a lack of explicit syntactic structures. While traditional masked language modeling (MLM) approaches do not explicitly focus on natural language interaction, emerging decoder-focused large language models (LLMs) demonstrate partial success in binary analysis yet remain underexplored for holistic comprehension. We present Assembly Augmented Tuning (\toolname), an end-to-end structural-semantic instruction tuning framework that synergizes encoder architecture with decoder-based LLMs through a projector module, where the assembly encoder extracts hardware-level structural features, the projector bridges representations with the semantic space, and the instruction-tuned LLM preserves natural language capabilities. Experimental results demonstrate three key advantages: (1) State-of-the-art performance in assembly comprehension with +39.7\% Recall@1 and +17.8\% MRR improvements over GPT-4-Turbo, (2) Consistent enhancements across base models (24.6-107.4\% Recall@1 and 15.2-106.3\% MRR on Qwen2.5-Coder, Deepseek-Coder and CodeLlama variants), and (3) Superior instruction-following capabilities (41.5\%-118\% improvements) with controlled code generation degradation (-8.9\% to -35\% across architectures).

\end{abstract}

%% file: sec/1_intro.tex
\section{Introduction}
The analysis and comprehension of assembly code play a pivotal role in critical domains such as reverse engineering~\citep{kargen2017towards, megira2018malware}, vulnerability detection~\citep{mantovani2022convergence, taviss2024asm2seq}, and software optimization~\citep{thirumoorthy2022feature}. Unlike high-level programming languages, assembly code lacks explicit syntactic structures and semantic abstractions, manifesting as low-information-density sequences of hardware-level instructions~\citep{wang2022jtrans}. This inherent complexity poses significant challenges for analysis, particularly when bridging assembly code semantics with natural language descriptions for downstream tasks like decompilation~\citep{tan2024llm4decompile}, code summarization~\citep{wang2021codet5}, and vulnerability explanation~\citep{al2023extending, xiong2023hext5}.
With the growing demand for assembly code comprehension, VirusTotal~\citep{VT-2023-05-12}, Google Cloud~\citep{Introducing-2023-04-25}, and Microsoft Security Copilot~\citep{Microsoft-2025-05-06} are increasingly leveraging LLMs to understand assembly code, aiming to provide powerful assistance to security professionals.

Existing approaches to assembly code comprehension primarily follow two paradigms. Methods based on masked language modeling (MLM)~\citep{su2024codeart, wang2024clap, ding2019asm2vec} excel at semantic extraction but are not explicitly focused on natural language interaction tasks.
Conversely, decoder-focused large language models (LLMs)~\citep{achiam2023gpt, touvron2023llama, chowdhery2023palm} demonstrate strong generation performance in source code domains and have powerful potential in binary code field~\citep{jin2023binary, shang2024far}.  
Recent works such as LLMCompiler~\citep{kim2023llm}, LLM4Decompile~\citep{tan2024llm4decompile}, and Nova~\citep{jiang2023nova} have verified the feasibility of LLMs through model training and domain adaptation. However, conventional approaches process assembly code through unified textual representations, failing to address the gap between hardware-centric assembly constructs and natural language semantics.

\begin{figure*}[tp]
    \centering
    \includegraphics[width=1\linewidth]{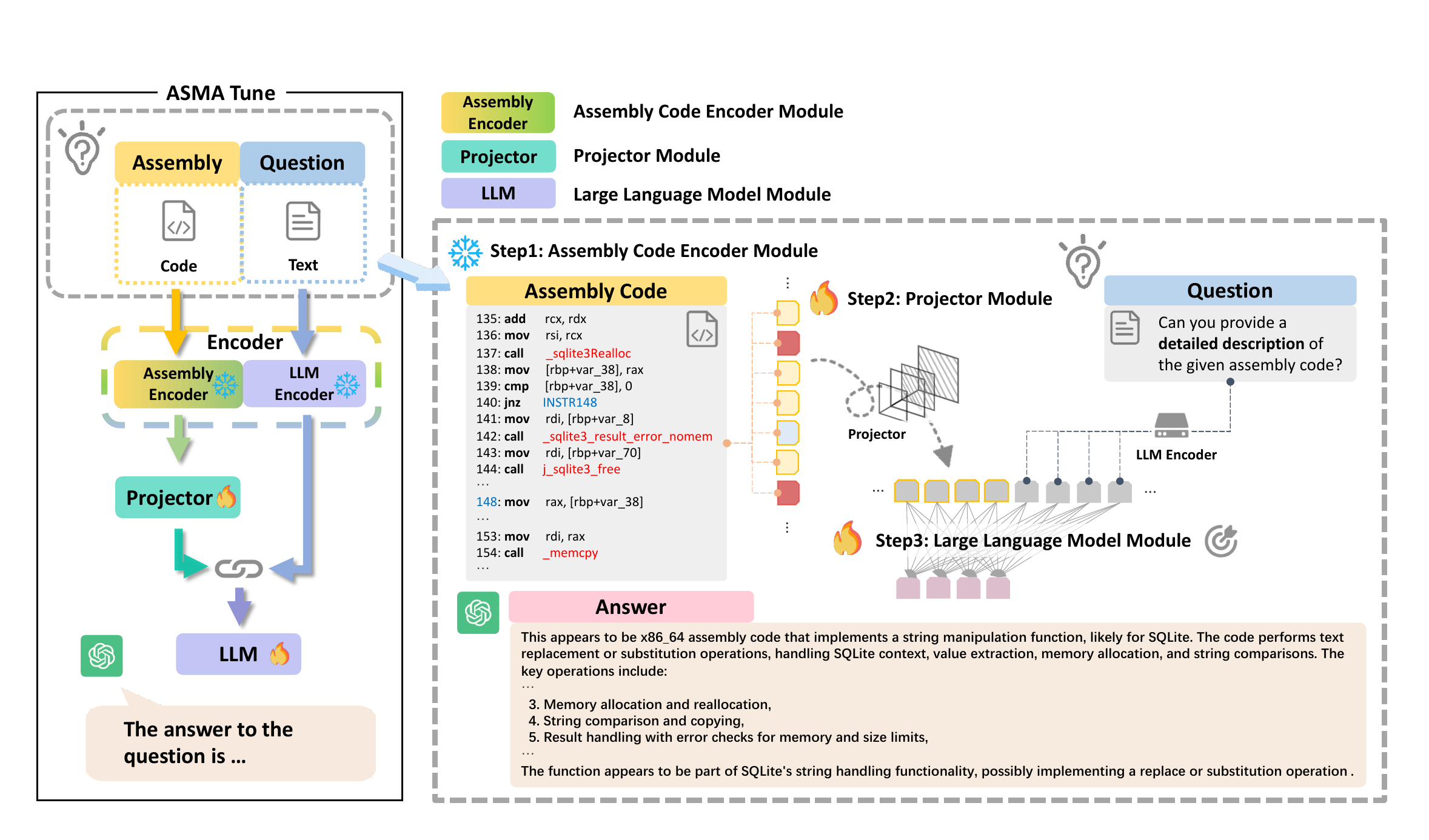}
    \caption{Overview of \toolname. It consists of three sub-modules: assembly code encoder, projector, and large language model. 
    }
    \vspace{2mm}
    \label{fig:overview}
\end{figure*}

To address these limitations, we introduce \textbf{A}s\textbf{S}e\textbf{M}bly \textbf{A}ugmented \textbf{T}uning (\toolname), the first end-to-end instruction tuning framework for joint assembly-language understanding and natural language interaction. Our framework enhances LLMs' ability to capture implicit logical dependencies in hardware-centric contexts and underlying assembly semantics that differ from natural language learning paradigms. 
As shown in Fig. \ref{fig:overview}, unlike directly feeding assembly code into an LLM, \toolname employs three core sub-modules: (1) \textit{Assembly Code Encoder Module} capturing hardware-level structural patterns, such as assembly vocabulary, control flow and external function names; (2) \textit{Projector Module} aligning assembly embeddings with LLM's semantic space; (3) \textit{LLM Module} maintaining natural language interactive capabilities. 

Our approach synergistically combines assembly-specific semantic grounding with LLM generation through structural-semantic instruction tuning. This integration enables seamless alignment between low-level assembly constructs and high-level textual descriptions, achieving superior performance over both general-purpose LLMs (e.g., GPT-4-Turbo~\citep{achiam2023gpt}) and code-specific LLMs (e.g., DeepSeek-Coder-V2-Lite-Instruct~\citep{zhu2024deepseek}).

Our main contributions are summarized as follows.

\begin{itemize}
\item We introduce \toolname, the end-to-end framework that aligns assembly features with natural language embeddings via a projector, connecting assembly-specific encoders with decoder-based LLMs. This architecture enhances both assembly code understanding and natural language instruction following capabilities.

\item We enable the creation of high-quality assembly instruction-following data by developing an instruction-centric generation pipeline. Additionally, we present ASMA-Bench, a comprehensive assembly-language instruction-following benchmark.

\item Our experiments demonstrate state-of-the-art performance across seven benchmarks and ASMA-Bench, achieving up to +39.7\% Recall@1 and +17.8\% MRR improvements over GPT-4-Turbo, with individual models reaching up to +107.4\% Recall@1 and +106.3\% MRR, as well as up to 118\% improvement in instruction-following capabilities. Furthermore, we will open-source our toolkit and 407K assembly-text instruction dataset once accepted to support future research.


\end{itemize}
\begin{figure}[t]
    \centering
    \includegraphics[width=1\linewidth]{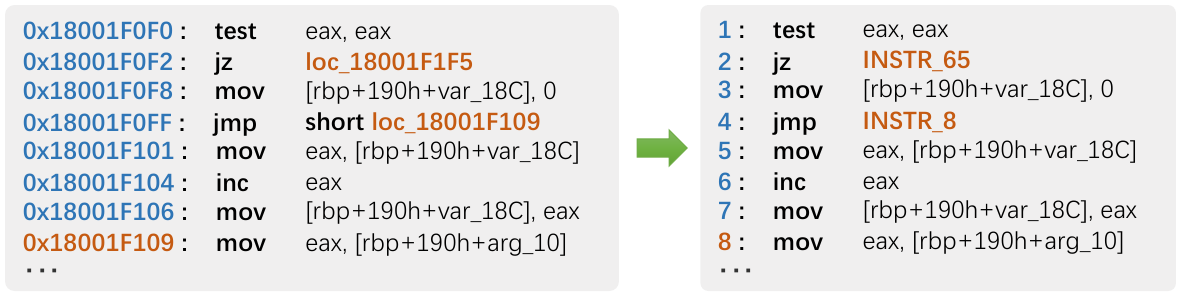}
    \caption{Instruction preprocess. 
    }
    \vspace{6mm}
    \label{fig:Instruction_preprocess}
\end{figure}

%% file: sec/2_related_work.tex
\section{Related Work}


\subsection{Assembly Code Representation and Analysis}

When high-level languages are transformed into assembly code, it is more difficult to understand and analyze because it is often accompanied by missing clear logical structures at the source code level~\citep{chlipala2011mostly}. Existing works on assembly code representation and analysis can be broadly categorized into models based on masked language modeling (MLM) and decoder-only LLMs.

MLM-based models excel in assembly code understanding tasks, although they are not explicitly focused on evaluation tasks involving natural language descriptions of binary code semantics. For instance, CodeArt~\citep{su2024codeart} introduces regularized attention masks to capture instruction semantics. CLAP~\citep{wang2024clap} improves analysis abilities in low-resource scenarios through natural language supervision of assembly-text pairs. Asm2Vec~\citep{ding2019asm2vec} further enables interpretable semantics of assembly code.

Decoder-focused models dedicated to improve the ability of large language models, such as GPT series~\citep{achiam2023gpt, floridi2020gpt}, LLaMA series~\citep{touvron2023llama, roziere2023code}, and other LLMs~\citep{zhu2024deepseek,hui2024qwen2}, which have shown strong capabilities in the understanding and summarization domain of source code, but their performance to discern semantic nuances in low-information-density assembly code still has substantial room for improvement~\citep{jin2023binary}. Nova~\citep{jiang2023nova} built on DeepSeek-Coder~\citep{guo2024deepseek}, employs hierarchical attention mechanisms and contrastive learning for joint understanding and generation tasks. Meta LLMCompiler trains CodeLlama to optimize binary code~\citep{cummins2024meta}. LLM4Decompile fine-tunes DeepSeekCoder for decompilation using assembly~\citep{tan2024llm4decompile}. 

In comparison, our method integrates an assembly code encoder with LLMs' natural language capabilities through end-to-end training, establishing a unified framework that bridges assembly-language processing and textual task execution.

\vspace{-2mm}

\subsection{Domain Specific Instruction Tuning}
\vspace{-2mm}

LLM instruction-tuning works such as InstructGPT~\citep{ouyang2022training}, FLAN-T5~\citep{chung2024scaling} and Vicuna~\citep{chiang2023vicuna}, have proven effective in enhancing the zero- and few-shot generalization of LLMs. In other communities, some studies use instruction tuning to align multiple modalities through end-to-end training or modular pipelines, such as InstructPix2Pix~\citep{brooks2023instructpix2pix}, LLaVA~\citep{liu2024visual} and InstructBLIP~\citep{panagopoulou2023x}.


However, applying instruction tuning to assembly language analysis remains thoroughly unexplored~\citep{jin2023binary}. We aim to address this gap and study its effectiveness by introducing \toolname, the first structural-semantic instruction-tuning framework for assembly-language understanding and instruction following capabilities. 

%% file: sec/3_method.tex
\vspace{-2mm}

\section{Methodology}

We present ASMA-Tune, a structural-semantic instruction tuning framework designed explicitly for assembly-language understanding. As illustrated in Fig.~\ref{fig:overview}, our solution operates through two coordinated phases: (1) assembly-language feature space alignment during pre-training, and (2) joint representation learning via end-to-end fine-tuning. This dual-phase architecture enhances semantic fidelity and analytical precision while preserving foundation models' generative capabilities. Subsequent sections detail our dataset engine pipeline, framework structure, and phased training methodology.

\vspace{-2mm}

\subsection{Dataset Engine} \label{sec:Dataset Engine}
The dataset engine is designed to systematically construct large-scale, diverse binary code datasets with semantic annotations and instruction-centric representations.

\vspace{-2mm}

\subsubsection{Binary Dataset Construction} \label{sec:Binary dataset construction}

\smallskip
\noindent
\textbf{Raw Dataset Selection.}
Previous research primarily relies on small-scale general-purpose codebases like GNU Utils and Coreutils to construct datasets, which exhibit convergent code patterns, lack diversity, and may lead to model overfitting. To overcome these limitations, we construct a multi-source dataset using BinaryCorp-3M~\citep{wang2022jtrans} and Juliet Test Suite~\citep{boland2012juliet}. 

BinaryCorp-3M is extracted from Arch Linux official repositories~\citep{Arch-Linux-Package-Search} and Arch User Repository(AUR) ~\citep{AUR-(en)-Home}, containing 10,265 binaries with 3M functions spanning diverse categories including editors, instant messaging tools, HTTP servers, web browsers, compilers, graphic libraries, and cryptographic libraries. It represents the largest and most diverse binary dataset for BCSD tasks to date~\citep{wang2022jtrans}, supporting both gcc/clang compilers with five optimization levels. Juliet Test Suite is a mature vulnerability analysis dataset containing 64,099 test cases with vulnerability descriptions. Its semantic annotations have been successfully applied to code comprehension tasks~\citep{taviss2024asm2seq}.

\smallskip
\noindent
\textbf{Binary Compilation and Disassembly.}
For the BinaryCorp-3M Train dataset, we reconstruct assembly instructions and their corresponding jump information in each function by parsing assembly instructions and control flow graphs using Capstone. Considering semantic complexity, we filtered functions with 10-3,000 instructions through random sampling across compilers and optimizations, yielding 212,117 instruction fragments from distinct functions. For the Juliet Test Suite, we compiled C/C++ source code with random gcc optimization levels (O0-O3) and disassembled them with IDAPro. Leveraging preserved symbol information from compilation and IDAPro's function identification capabilities, we parsed function names and types in binaries. We constructed function descriptions through Source-Sink combinations from vulnerability description headers in source code (including GoodSource, GoodSink, BadSource, and BadSink). After removing functions with identical descriptions, we finally obtained 79,920 annotated distinct assembly function samples.

\smallskip
\noindent
\textbf{Instruction Pre-processing.}
To effectively represent assembly code while capturing its inherent structure and semantics, and preserving critical information such as function call parameters, strings processed by decompilers, and variable names. Consistent with prior work~\citep{wang2024clap}, we rebase the addresses in assembly code, keeping relative address relationships when handling jump instructions. This ensures that intra-function jump instructions maintain their relative distances to targets, preserving control flow information and enabling the model to comprehend the control flow changes induced by jump instructions. Fig.~\ref{fig:Instruction_preprocess} is a processing example.



\subsubsection{Instruction-Centric Syntheic Data Generation}\label{sec:Instruction-Centric Data Generation}

Our assembly code corpus integrates 292,037 assembly snippets, 212,117 from BinaryCrop-3M Train and 79,920 from Juliet Test Suite. Next, we construct two datasets comprising four types of conversation generated through LLM. Given an assembly code corpus \( C = \{c_i\}_{i=1}^N \), we formulate the instruction data generation process as follows.
\begin{equation}
\small
\mathrm{G}(c, t) = \mathcal{M}_\phi\big(\pi_t(c)\big) \mapsto (Q_t, A_t)
\label{equation:Generation}
\end{equation}

Where \( t \in T = \{\text{simp}, \text{detail}, \text{conv}, \text{reason}\} \) denotes the task type, \(Q_t\) represents the generated question for type \(t \), \(A_t\) denotes the corresponding target answer, and \( \pi_t: C \to P \) represents task-specific prompt engineering with few-shot demonstrations.

Specifically, for assembly code $c$, we employ $\mathcal{M}_{\phi}$ to generate responses under task-specific instructions. Through carefully designed prompts integrating few-shot examples and in-context learning, model produces corresponding outputs for each instruction type.

We have curated four types of supervised instruction-following training data: simplified description ${q}_{simp}$, detailed description ${q}_{detail}$, multiturn conversations ${q}_{conv}$, and complex reasoning ${q}_{reason}$. This results in two trained datasets, which are defined as follows.

\begin{equation}
\small
\mathrm{D}_1^* = \bigcup_{c \in C} \left\{ \mathrm{G}(c, \text{simp}) \oplus c \right\}
\label{equation:D1}
\end{equation}

\begin{equation}
\small 
\mathrm{D}_2^* = \bigcup_{c \in C} \left\{ \mathrm{G}(c, t) \oplus c \mid t \in T \{\text{detail, conv, reason}\} \right\}
\label{equation:D2}
\end{equation}

$\mathrm{D}_1^*$ serves for pre-training and $\mathrm{D}_2^*$ for fine-tuning. The instruction type distribution in $\mathrm{D}_2^*$ follows~\citep{liu2024visual}. The constructed dataset $\mathrm{D}=\{\mathrm{D}_1^*, \mathrm{D}_2^*\}$ provides multi-granularity supervision signals for training. 
Complete training dataset details are presented in Tab. \ref{tab:datasets_description}.

\input{tables/datasets_description}


\subsection{Framework Structure}  \label{sec:Model Structure}

As illustrated in Fig. \ref{fig:overview}, the \toolname framework comprises three key components: assembly code encoder $\mathcal{F}_{c}$ extracts structural-semantic features from input code, projector $\mathcal{W}$ bridges the representational gap between assembly features and natural language embeddings, and pretrained large language model $\mathcal{M}_{\phi}$ enables instruction-guided semantic understanding and generation.

The architecture employs several special tokens to help with assembly code processing.
\begin{itemize}
\item \verb$<|inst_start|>$ and \verb$<|inst_end|>$ establish explicit instruction boundaries.
\item \verb$<|inst_code|>$ encapsulates assembly snippets to preserve code semantics.
\item \verb$<|inst_placeholder|>$ enables code-text alignment.
\end{itemize}

The dimension-aligned projector $\mathcal{W}$ implements feature space transformation through learnable linear layers.

\begin{equation}
\small
\mathcal{W}: \mathbb{R}^{d_{enc}} \rightarrow \mathbb{R}^{d_{llm}}
\end{equation}

Where $d_{enc}$ denotes the assembly encoder's output dimension and $d_{llm}$ the LLM's embedding dimension. \textit{This projection enables cross-modal representation alignment while preserving architectural flexibility for different encoder-LLM pairs. Alg. \ref{algorithm1} outlines the specifics of \toolname interactive assembly understanding. }

\begin{algorithm}[t]
\caption{Structural-Semantic Instruction Tuning}\label{algorithm1}
\SetKwInput{KwInput}{Input}                
\SetKwInput{KwOutput}{Output}              
\DontPrintSemicolon
  \KwInput{Assembly code $c$, Question set $\mathrm Q$, Assembly encoder $\mathcal{F}_{c}$, Pre-trained projector $\mathcal{W}$, LLM $\mathcal{M}_{\phi}$}
  \KwOutput{Answer set $\mathrm A$}
  
  \SetKwFunction{FEncode}{AssemblyEncoder}
  \SetKwFunction{FGenerate}{LLM\_Generate}
  \SetKwProg{Fn}{Function}{:}{}

  \Fn{\FEncode{$c$}}{
    $c_{\text{wrap}} \leftarrow \texttt{<|inst\_start|>} \oplus c \oplus \texttt{<|inst\_end|>}$\;
    ${E}_c \leftarrow \mathcal{F}_c(c_{\text{wrap}})$\;
    $\tilde{{E}}_c \leftarrow \mathcal{W}({E}_c) \oplus \texttt{<|inst\_placeholder|>}$\;
    \KwRet $\tilde{{E}}_c$\;
  }
  
  \Fn{\FGenerate{$c$, $\mathrm Q$}}{
    $\tilde{{E}}_c = \texttt{AssemblyEncoder}(c)$\;
    $\mathrm A \leftarrow \emptyset$\;
    \ForEach{$q \in \mathrm Q$}{
      ${E}_q \leftarrow \mathcal{M}_{\phi}.\text{Embed}(\texttt{<|inst\_code|>} \oplus q)$\;

      ${E}_t = \tilde{{E}}_c \oplus {E}_q$\;
      $a \leftarrow \mathcal{M}_\phi.\text{Generate}({E}_t)$\;
      $\mathrm{A} \leftarrow \mathrm{A} \cup \{a\}$\;
    }
    \KwRet $\mathrm A$\;
  }
  
  $\mathrm A \leftarrow \texttt{LLM\_Generate}(c, \mathrm Q)$\;
\end{algorithm}


With the aligned feature space established by $\mathcal{F}_{c}$ and $\mathcal{W}$, we design a two-phase training paradigm to optimize the model's assembly-language understanding capability progressively.


\subsection{Training Methodology} \label{sec:Training Methodology}

\subsubsection{Feature Space Alignment Pre-training} \label{sec:Pre-training}
We initialize the projector $\mathcal{W}$ and train it on $\mathrm {D_{1}^{*}}$ (refer to Eq. \ref{equation:D1}) to align the assembly code features with their corresponding simplified descriptions.

For an input assembly code ${c}$ and its associated simplified query ${q}_{simp}$ and answer ${a}_{simp}$, we utilize the assembly encoder $\mathcal{F}_{c}$ to obtain its embedding ${E}_{c} = \mathcal{F}_{c}(c)$. The projector processes the assembly grid features before and after the final transformer layer, aligning them into the language word embedding space. After transformation by the projector $\mathcal{W}$, we obtain the assembly code feature in the language word embedding space as: ${E}_{c}\cdot \mathcal{W}$. We represent this process using an auto-regressive training loss, defined as:

\begin{equation}
\small \mathcal{L}_{\text{pretrain}} = \mathbb{E}{(Q,A,c) \sim \mathrm{D}_1^*}\left[- \sum_{i=1}^{L}\log P_{\theta} \left( A_i \big| \mathrm{D}_{<i}^* \right) \right]
\label{equation:projector loss}
\end{equation}

Where \(Q\) denotes the input question, \(A\) represents the target answer, \(c\) is the input assembly code,  ${L}$ is the sequence length, $\mathrm {A}_{i}$ is the $i$-th token of the target answer, $\mathrm {D}_{< i}^{*}$ represents the tokens from all previous turns before the current prediction tokens, $\theta = \left \{\mathcal{W}\right \}$ denotes the trainable parameters, and ${P}\left( \mathrm {A}_{i}| \mathrm {D}_{< i}^{*};\theta\right)$ is the probability that the model predicts $\mathrm {A}_{i}$ given $\mathrm {D}_{< i}^{*}$.

Through the first stage of projector pre-training, we obtain a well-initialized set of weights for the subsequent fine-tuning stage, accelerating the convergence of the fine-tuning process.


\subsubsection{End-to-End Fine-tuning} \label{sec:Fine-tuning}

This section explains how the \toolname framework fine-tunes and predicts answers for assembly code.

We continue updating both the pre-trained weights of the projection layer and the LLM on the dataset $\mathrm {D_{2}^{*}}$ (refer to Eq. \ref{equation:D2}). The updated auto-regressive training loss is defined as:

\begin{equation}
\small
\mathcal{L}_{\text{finetune}} = \mathbb{E}{(Q,A,c) \sim \mathrm{D}_2^*} \left[- \sum_{i=1}^{L}\log P_{\theta} \left( A_i \big| \mathrm{D}_{<i}^* \right) \right]
\label{equation:LLM loss}
\end{equation}

$\theta = \left \{\mathcal{W,M}\right \}$ encompasses the trainable parameters, including the projection matrix $\mathcal{W}$ and the LLM parameters $\mathcal{M}_{\phi}$.

By renewing this auto-regressive loss, we enhance the mutual information between assembly code and natural language, effectively aligning their representations. This alignment facilitates more accurate and meaningful predictions of answers for assembly code.

%% file: tables/datasets_description.tex
\begin{table}[t]
  \centering
  \caption{Training dataset composition and sources.}

  \begin{threeparttable}
    \begin{tabular}{l|p{12em}|p{4em}}
    \toprule
    \textbf{Dataset} & \multicolumn{1}{l|}{\textbf{Description}} & \textbf{Record} \\
    \midrule
    \multicolumn{3}{l}{\textbf{Raw Data}} \\
    \midrule
    BinaryCorp-3M Train & 8,357 binaries, O0-O4 optimizations, gcc/clang compilers. & 212,117 \\
    \midrule
    Juliet Test Suite & C/C++ vulnerability test cases with function pairs. & 79,920 \\
    \midrule
    \multicolumn{3}{l}{\textbf{Syntheic Data}} \\
    \midrule
    $\mathrm{D}_1^*$$^\dagger$(Pre-training) & Simplified assembly-text QA pairs. & 292,037 \\
    \midrule
    $\mathrm{D}_2^*$$^\dagger$(Fine-tuning) & Detailed descriptions (28k), conversations (43k), reasoning QA (44k). & 115,034 \\
    \bottomrule
    \end{tabular}%
    \begin{tablenotes}[para,flushleft]
        \item $^\dagger$ Generated using GPT-4-Turbo.
    \end{tablenotes}
  \end{threeparttable}
  \label{tab:datasets_description}%

\end{table}%

%% file: sec/4_experiments.tex
\section{Experimental Setup}

\subsection{Evaluation Datasets}
\smallskip
\noindent
\textbf{Binary Code Similarity Detection (BCSD).} We evaluate on seven established benchmarks (Curl, Coreutils, Binutils, ImageMagick, SQLite, OpenSSL, Putty) widely used in binary semantics researches~\citep{pei2020trex, su2024codeart, jiang2023nova}. 

\smallskip
\noindent
\textbf{Assembly-Language Instruction-Following.} We present \textbf{ASMA-Bench}, \textit{the first benchmark for instruction-following in assembly language.} 30 domain-diverse assembly snippets are randomly selected from the BinaryCrop-3M Test dataset (containing 364 projects and 1,908 binaries with non-overlapping test samples excluded from training data), with 3 questions per snippet systematically generated through the pipeline in Section~\ref{sec:Instruction-Centric Data Generation}, yielding a 90-question evaluation suite. The benchmark spans multiple domains including cryptography, malware, and protocol implementations, comprising three distinct question types: conversation, description, and reasoning.

\smallskip
\noindent
\textbf{Code Generation.} HumanEval~\citep{chen2021evaluating} benchmark consists of 164 hand-written Python programming tasks, each providing a function signature and docstring as model input. Widely adopted for evaluating code LLMs~\citep{guo2024deepseek, hui2024qwen2} in a zero-shot setting.




\begin{figure*}
    \centering 
    \includegraphics[width=1\linewidth]{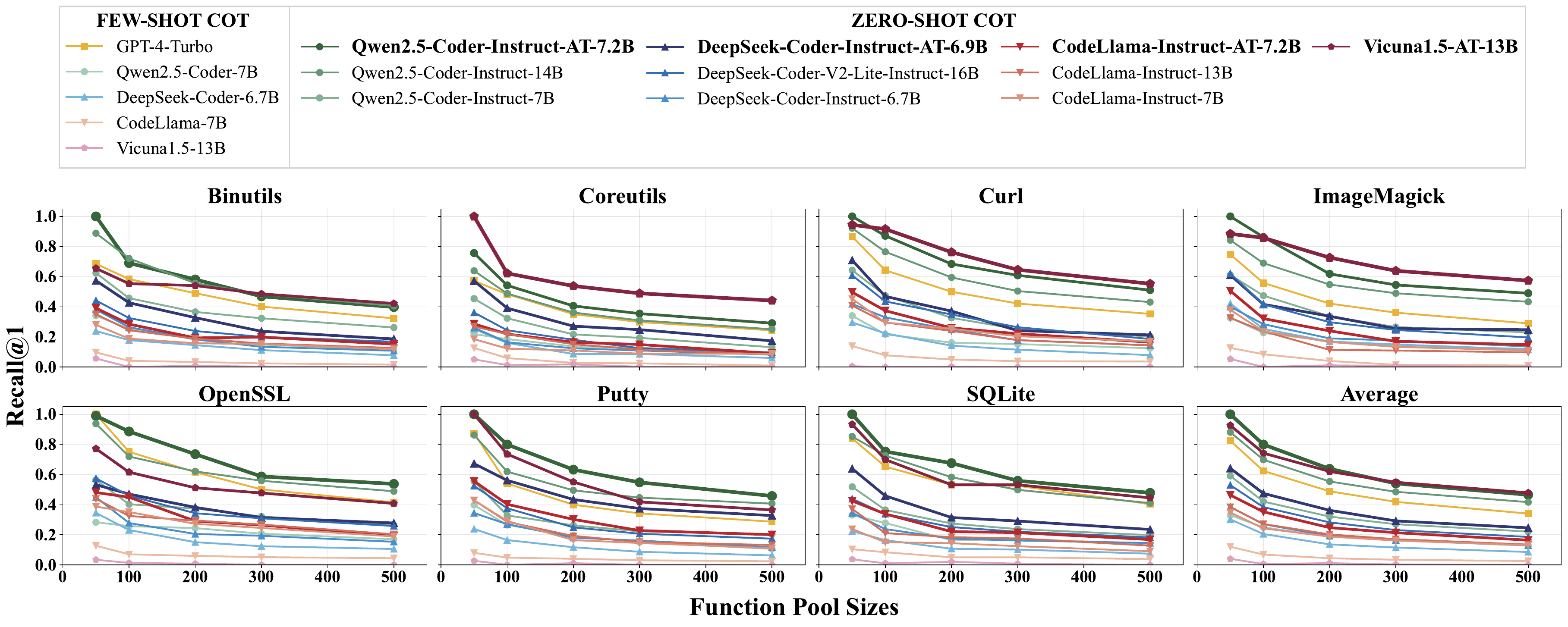}
    \caption{BCSD experimental results. \toolname enhanced models inherit base architecture identifiers with -\textbf{AT} suffix. We use normalized evaluation results across datasets. The x-axes denote different sizes of candidate function pools, and the y-axes denote the Recall@1 performance.}
    \vspace{2mm}

    \label{fig:BCSD_recall}
\end{figure*}

\subsection{Models} \label{sec:models}

Our method comprises three distinct components: \textbf{Assembly Code Encoder} employing CLAP-ASM from~\citep{wang2024clap} with 110M parameters generating 768-dimensional embeddings, \textbf{Projector} initialized through LLaVA's efficient alignment methodology by ~\citep{liu2024visual}, implementing a single-layer MLP transformation $\mathbb{R}^{768} \rightarrow \mathbb{R}^{d_{llm}}$ with 30M params, and \textbf{Large Language Model}. Dimension alignment adapts each LLM's hidden size, 3,584 for Qwen2.5-Coder-Instruct-7B, 4,096 for DeepSeek-Coder-Instruct-6.7B/CodeLlama-Instruct-7B, and 5,120 for Vicuna1.5-13B.

\subsection{Training} \label{sec:training}


Training executes on 4 NVIDIA Tesla A100 80G GPUs and Pytorch framework through two strategic phases with BF16 and DeepSpeed ZeRO-2 optimization.

\smallskip
\noindent
\textbf{Aligned Pre-training.}
The projector trains via Eq. ~\ref{equation:projector loss}'s auto-regressive objective on $\mathrm {D_{1}^{*}}$ (292k samples) dataset, completing one epoch in 2 hours with a 2e-3 learning rate and 128 batch size.

\smallskip
\noindent
\textbf{Fine-tuning.}
End-to-end fine-tuning via Eq. ~\ref{equation:LLM loss} executes for three epochs totaling 6 hours on $\mathrm {D_{2}^{*}}$ (115k samples) dataset, employing a 2e-5 learning rate and 32 batch size.

\subsection{Baselines} \label{sec:Baselines}


We evaluate state-of-the-art (SOTA) open-source code instruct models with comparable parameter scales (6.7B-16B parameters) and the mainstream LLMs under identical experimental settings.

\smallskip
\noindent
\textbf{GPT-4-Turbo}~\citep{achiam2023gpt} developed by OpenAI, stands out as one of the top-performing models in code capability assessments within the code benchmark~\citep{zhuo2024bigcodebench}, showcasing unparalleled proficiency in code understanding tasks.

\smallskip
\noindent
\textbf{Vicuna1.5}~\citep{chiang2023vicuna} is a general-purpose chat assistant fine-tuned from Llama 2, serving as a representative baseline for non-code-specialized LLMs.

\smallskip
\noindent
\textbf{CodeLlama-Instruct}~\citep{roziere2023code} is an instruction-tuned generative text model specifically fine-tuned for general code synthesis and understanding.

\smallskip
\noindent
\textbf{DeepSeek-Coder-V2-Instruct}~\citep{zhu2024deepseek} emerges as a cutting-edge Mixture-of-Experts model trained on 6 trillion tokens, achieving GPT-4-Turbo level performance in code-specific evaluations.

\smallskip
\noindent
\textbf{Qwen2.5-Coder-Instruct}~\citep{hui2024qwen2} offers long-context support of up to 128K tokens for Code Agents.

Our method enhances existing instruction-tuned code models through \toolname while maintaining comparisons with both their original instruction-tuned versions (e.g., CodeLlama-Instruct) and corresponding base models (e.g., CodeLlama).



\input{tables/eval-BCSD_MRR}

\input{tables/eval-ASMABench}


\vspace{-2mm}

\subsection{Metrics} \label{sec:Metrics}

\smallskip
\noindent
\textbf{BCSD Evaluation.}
We employ two standard metrics consistent with prior research~\citep{su2024codeart, jiang2023nova, wang2024clap}. Recall@1 is the percentage of queries where the ground-truth candidate ranks first, and Mean Reciprocal Rank (MRR) measures the average reciprocal rank of the correct candidate across all queries.

\smallskip
\noindent
\textbf{ASMA-Bench Evaluation.}
GPT-4-Turbo establishes a performance upper bound through reference predictions generated using ground-truth textual descriptions. The Evaluation Arbitrator (GPT-4-Turbo) scores candidate model outputs relative to the GPT-4-Turbo baseline on a scale from 1 to 10 across five dimensions: helpfulness, relevance, accuracy, detail, and comprehensiveness. Higher scores reflect superior performance.


\smallskip
\noindent
\textbf{HumanEval Evaluation.}
We adopt the standard Pass@1 metric to measure the proportion of functionally correct code generations that pass all test cases, with execution outcomes rigorously validated through automated testing (pass/fail/timeout states) in zero-shot settings.

\vspace{-2mm}

\subsection{Implementation Details} \label{sec:Implementation Details}

Our BCSD evaluation employs semantic retrieval comparison: all methods generate assembly explanations using identical prompts to build retrieval corpora. Input descriptions are created with the same template and using Contriever-MSMARCO~\citep{izacard2021unsupervised} as the retriever. Following ~\citep{ram2023context}, retrieval length $l = 64$ balances efficiency and information richness. Inference uses a temperature of 0.2, top-k sampling set to 1, maximum input length constrained by the model's limit and maximum output length of 512 tokens. All experiments prevent data leakage by excluding prompts during corpus generation.

%% file: tables/eval-BCSD_MRR.tex
\begin{table*}[tbp]
  \centering

  \caption{BCSD experimental results. Bold number indicates the best performance.}
  \setlength{\tabcolsep}{2pt}
    \begin{tabular}{lr|cccccccc}
    \toprule
    \multirow{2}[2]{*}{\textbf{Model}} & \multicolumn{1}{c|}{\multirow{2}[2]{*}{\textbf{Size}}} & \multicolumn{8}{c}{\textbf{BCSD Evalution (Pool size 500)}} \\
\cmidrule{3-10}          &       & \multicolumn{1}{c}{\textbf{Binutils}} & \multicolumn{1}{c}{\textbf{Coreutils}} & \multicolumn{1}{c}{\textbf{Curl}} & \multicolumn{1}{c}{\textbf{ImageMagick}} & \multicolumn{1}{c}{\textbf{SQLite}} & \multicolumn{1}{c}{\textbf{OpenSSL}} & \multicolumn{1}{c}{\textbf{Putty}} & \multicolumn{1}{c}{\textbf{Average}} \\
    \midrule
    \multicolumn{10}{c}{\textit{(FEW-SHOT COT)}} \\
    \midrule
    GPT-4-Turbo & -     & 0.190 & 0.175  & 0.191  & 0.162  & 0.169  & 0.232  & 0.174  & 0.185  \\
    \textcolor[rgb]{ .122,  .137,  .161}{Qwen2.5-Coder} & \textcolor[rgb]{ .122,  .137,  .161}{7B} & \textcolor[rgb]{ .122,  .137,  .161}{0.072} & \textcolor[rgb]{ .122,  .137,  .161}{0.076} & \textcolor[rgb]{ .122,  .137,  .161}{0.078} & \textcolor[rgb]{ .122,  .137,  .161}{0.072} & \textcolor[rgb]{ .122,  .137,  .161}{0.065} & \textcolor[rgb]{ .122,  .137,  .161}{0.101} & \textcolor[rgb]{ .122,  .137,  .161}{0.070} & \textcolor[rgb]{ .122,  .137,  .161}{0.076} \\
    DeepSeek-Coder & 6.7B  & 0.050  & 0.053  & 0.052  & 0.074  & 0.037  & 0.067  & 0.047  & 0.054  \\
    CodeLlama & 7B    & 0.017  & 0.019  & 0.024  & 0.018  & 0.020  & 0.027  & 0.019  & 0.021  \\
    Vicuna1.5 & 13B   & 0.006  & 0.010  & 0.007  & 0.009  & 0.005  & 0.008  & 0.005  & 0.007  \\
    \midrule
    \multicolumn{10}{c}{\textit{(ZERO-SHOT COT)}} \\
    \midrule
    Qwen2.5-Coder-Instruct & 14B   & \textbf{0.227} & 0.179  & 0.224  & 0.226  & 0.172  & 0.246  & 0.212  & 0.212  \\
    Qwen2.5-Coder-Instruct & 7B    & 0.155  & 0.106  & 0.119  & 0.135  & 0.091  & 0.147  & 0.119  & 0.124  \\
    \rowcolor[rgb]{ .906,  .902,  .902} \textbf{Qwen2.5-Coder-Instruct-AT} & \textbf{7.2B} & \textbf{0.225} & \textbf{0.194} & \textbf{0.244} & \textbf{0.245} & \textbf{0.189} & \textbf{0.268} & \textbf{0.239} & \textbf{0.257} \\
    \midrule
    DeepSeek-Coder-V2-Lite-Instruct & 16B   & 0.108  & 0.086  & 0.109  & 0.119  & 0.086  & 0.141  & 0.107  & 0.108  \\
    DeepSeek-Coder-Instruct & 6.7B  & 0.075  & 0.070  & 0.083  & 0.084  & 0.066  & 0.095  & 0.075  & 0.078  \\
    \rowcolor[rgb]{ .906,  .902,  .902} \textbf{DeepSeek-Coder-Instruct-AT} & \textbf{6.9B} & \textbf{0.114} & \textbf{0.121} & \textbf{0.109} & \textbf{0.136} & \textbf{0.101} & \textbf{0.152} & \textbf{0.163} & \textbf{0.128} \\
    \midrule
    CodeLlama-Instruct & 13B   & 0.079  & 0.068  & 0.083  & 0.066  & 0.062  & \textbf{0.119}  & 0.082  & 0.080  \\
    CodeLlama-Instruct & 7B    & 0.080  & 0.070  & \textbf{0.095}  & 0.078  & 0.046  & 0.113  & 0.071  & 0.079  \\
    \rowcolor[rgb]{ .906,  .902,  .902} \textbf{CodeLlama-Instruct-AT} & \textbf{7.2B} & \textbf{0.092} & \textbf{0.072} & \textbf{0.088} & \textbf{0.087} & \textbf{0.074} & \textbf{0.113} & \textbf{0.109} & \textbf{0.091} \\
    \midrule
    \rowcolor[rgb]{ .906,  .902,  .902} \textbf{Vicuna1.5-AT} & \textbf{13.2B} & \textbf{0.208} & \textbf{0.258} & \textbf{0.242} & \textbf{0.261} & \textbf{0.158} & \textbf{0.200} & \textbf{0.202} & \textbf{0.218} \\
    \bottomrule
    \end{tabular}%
  \label{tab:eval-BCSD(MRR)}%
\end{table*}%

%% file: tables/eval-ASMABench.tex
\begin{table*}[t]
  \centering
  \vspace{2mm}

  \caption{ASMA-Bench and HumanEval results. For ASMA-Bench, GPT-4-Turbo responses were collected through three independent query trials to ensure evaluation consistency.}


    \begin{tabular}{lr|cccc|c}
    \toprule
    \multirow{2}[2]{*}{\textbf{Model}} & \multicolumn{1}{c|}{\multirow{2}[2]{*}{\textbf{Size}}} & \multicolumn{4}{c|}{\textbf{ASMA-Bench}} & \textbf{HumanEval} \\
\cmidrule{3-7}          &       & Conversation & Detail description & Complex reasoning  & All   & Pass@1 \\
    \midrule
    \textcolor[rgb]{ .122,  .137,  .161}{GPT-4-Turbo} & \textcolor[rgb]{ .122,  .137,  .161}{-} & 82.65  & 85.42  & 87.20  & 85.09  & - \\
    Vicuna1.5 & 13B   & 12.30  & 15.60  & 11.00  & 12.97  & - \\

    \midrule
    Qwen2.5-Coder-Instruct & 14B   & 81.67  & 63.60  & 76.63  & 73.97  & 89.60  \\
    Qwen2.5-Coder-Instruct & 7B    & 72.30  & 56.10  & 62.61  & 63.67  & 88.40  \\
    \rowcolor[rgb]{ .906,  .902,  .902} \textbf{Qwen2.5-Coder-Instruct-AT} & \textbf{7.2B} & \textbf{88.11} & \textbf{92.52} & \textbf{90.00} & \textbf{90.21} & 80.50  \\
    \midrule
    DeepSeek-Coder-V2-Lite-Instruct & 16B   & 74.98  & 67.47  & 69.40  & 70.62  & 81.10  \\
    DeepSeek-Coder-Instruct & 6.7B  & 69.33  & 59.50  & 55.84  & 61.56  & 66.10  \\
    \rowcolor[rgb]{ .906,  .902,  .902} \textbf{DeepSeek-Coder-Instruct-AT} & \textbf{6.9B} & \textbf{87.10} & \textbf{91.62} & \textbf{82.67} & \textbf{87.13} & 61.00  \\
    \midrule
    CodeLlama-Instruct & 13B   & 48.62  & 33.88  & 52.73  & 45.08  & 42.70  \\
    CodeLlama-Instruct & 7B    & 35.40  & 26.67  & 30.00  & 30.69  & 34.80  \\
    \rowcolor[rgb]{ .906,  .902,  .902} \textbf{CodeLlama-Instruct-AT} & \textbf{7.2B} & \textbf{62.22} & \textbf{70.02} & \textbf{68.34} & \textbf{66.88} & 22.60  \\
    \midrule
    \rowcolor[rgb]{ .906,  .902,  .902} \textbf{Vicuna1.5-AT} & \textbf{13.2B} & \textbf{90.10} & \textbf{85.15} & \textbf{86.12} & \textbf{87.12} & \cellcolor[rgb]{ 1,  1,  1}- \\
    \bottomrule
    \end{tabular}%
  \label{tab:eval-ASMABench}%
\end{table*}%

%% file: sec/5_results.tex
\section{Experiment Results}

\subsection{Main Results 1: Binary Code Similarity Detection Evaluation}\label{Main result 1}
Our experimental findings are systematically presented in Fig. ~\ref{fig:BCSD_recall} and Tab. ~\ref{tab:eval-BCSD(MRR)}. The results demonstrate \toolname enhanced models' (denoted with \textbf{AT} suffix) Recall@1 performance across varying function pool sizes and their MRR metrics (evaluated with 500-function pools) compared to existing techniques on seven BCSD benchmarks. Key observations are as follows.

\smallskip
\noindent
\textbf{\toolname enhanced models consistently outperform their original versions and baseline methods across all 7 datasets}, achieving state-of-the-art average Recall@1 (0.367) and MRR (0.257) with 500-function pools. Specifically, the \toolname enhanced Vicuna1.5-\textbf{AT}-13.2B exhibits a +39.7\% relative improvement in Recall@1 and +17.8\% in MRR compared to GPT-4-Turbo under 500-function pool configurations. When compared to open-source code LLMs, \toolname enhanced models demonstrate superior assembly code comprehension capabilities. Qwen2.5-Coder-Instruct-\textbf{AT}-7.2B achieves +107.4\% higher Recall@1 and +106.3\% better MRR than its base model (Qwen2.5-Coder-Instruct-7B), while DeepSeek-Coder-Instruct-\textbf{AT}-6.9B shows +84.8\% improvement in Recall@1 and +64.1\% in MRR compared to DeepSeek-Coder-Instruct-6.7B, all measured with 500-function pools.

\smallskip
\noindent
\textbf{The performance gains are particularly significant for general-purpose LLMs}, where assembly understanding plays a critical role. For instance, Vicuna1.5-\textbf{AT}-13.2B achieves a 75× relative improvement in Recall@1 and a 30× enhancement in MRR compared to Vicuna1.5-13B under 500-function pool conditions. Across all datasets, \toolname enhanced models maintain an average 85.9\% higher MRR than non-enhanced counterparts in 500-function pool configurations. These results validate our method's effectiveness in enhancing models' capacity to discern subtle assembly patterns in complex scenarios, particularly at larger pool sizes where semantic understanding of low-level code becomes paramount.


\input{tables/Ablation}
\vspace{-2mm}

\subsection{Main Results 2: Assembly-Language Instruction Following Evaluation}\label{Main result 2}
The evaluation results on ASMA-Bench benchmark are presented in Tab. \ref{tab:eval-ASMABench}, assessing instruction following capabilities and assembly code comprehension in novel domains.

\smallskip
\noindent
\textbf{Models enhanced with \toolname demonstrate superior performance in conversation, detailed description, and complex reasoning tasks.}

Experimental results show that \toolname enhanced models achieve state-of-the-art performance, with Qwen2.5-Coder-Instruct-\textbf{AT}-7.2B attaining an average score of 90.21, +41.7\% improvement over its non-enhanced version. DeepSeek-Coder-Instruct-\textbf{AT}-6.9B outperforms its base model by +41.5\%, while CodeLlama-Instruct-\textbf{AT}-7.2B demonstrates a +118\% improvement. Notably, \toolname enhanced models matching or even surpassing GPT-4-Turbo's performance (85.09), with Vicuna1.5-\textbf{AT}-13.2B achieving comparable results (87.12) despite having 10× fewer parameters, underscoring the effectiveness of \toolname in instruction following.

\vspace{-2mm}

\subsection{Main Results 3: General Code Generation Capability Preservation.}\label{Main result 3}
To verify the preservation of fundamental general code generation capabilities, we evaluate models on the HumanEval benchmark. Results are presented in Tab. \ref{tab:eval-ASMABench}.

\smallskip
\noindent
\textbf{\toolname enhanced models maintain competitive code generation performance with controllable degradation.} Experimental results show that \textbf{AT} models retain controlled original capabilities trade-off. Qwen2.5-Coder-Instruct-\textbf{AT}-7.2B achieves 80.5 Pass@1 (-8.9\%) and DeepSeek-Coder-Instruct-\textbf{AT}-6.9B scores 61 (-7.7\%),while CodeLlama-Instruct-\textbf{AT}-7.2B decreased by 35\%. This validates that \toolname effectively preserves models' general coding proficiency while specializing them for assembly-language understanding.

\vspace{-2mm}

\subsection{Ablation Study}\label{Ablation Study}
We conduct comprehensive ablation studies to quantify the contribution of each component in \toolname, using Vicuna1.5-\textbf{AT}-13.2B as a case study. Ablation studies validate that the synergistic interaction between our model's core architectural components and domain-specific training data collectively drives superior performance.

\vspace{-2mm}

\subsubsection{Training Data Ablation}
We systematically removed Detailed description, Conversation, and Complex reasoning data from the training set and evaluated on the ASMA-Bench dataset. Results are shown in Tab. \ref{tab:ablation1}.

\smallskip
\noindent
\textbf{All three instruction types are essential for optimal assembly comprehension.} The full-data configuration achieves peak performance (87.12), while removing any data component degrades results. Notably, excluding conversation data leads to the most severe overall performance drop (-46.31). Removing detailed description impacts description tasks (-68.2), suggesting these data anchor the model's ability to produce accurate analyses. The absence of complex reasoning data severely degrades complex reasoning performance (-75.2), confirming its essential role in multi-step problem solving.
\vspace{-2mm}

\subsubsection{Model Training and Size Ablation}
Tab. \ref{tab:ablation2} demonstrates the impact of architectural choices and scale on BCSD task performance measured by average Recall@1.

\smallskip
\noindent
\textbf{Complete training pipeline and model scale synergistically enhance performance.}

The assembly encoder contributes 52\% of post-training gains, while the remaining 48\% stems from the LLM's intrinsic reasoning. Omitting pre-training (-38.84) or fine-tuning (-43.29) severely degrades performance, validating their complementary roles in structural-semantic alignment. The 13.2B model outperforms its 7B counterpart by 5.04 Recall@1 points, demonstrating the benefits of scale.


\begin{figure}[ht]
    \centering 
    \includegraphics[width=.5\linewidth]{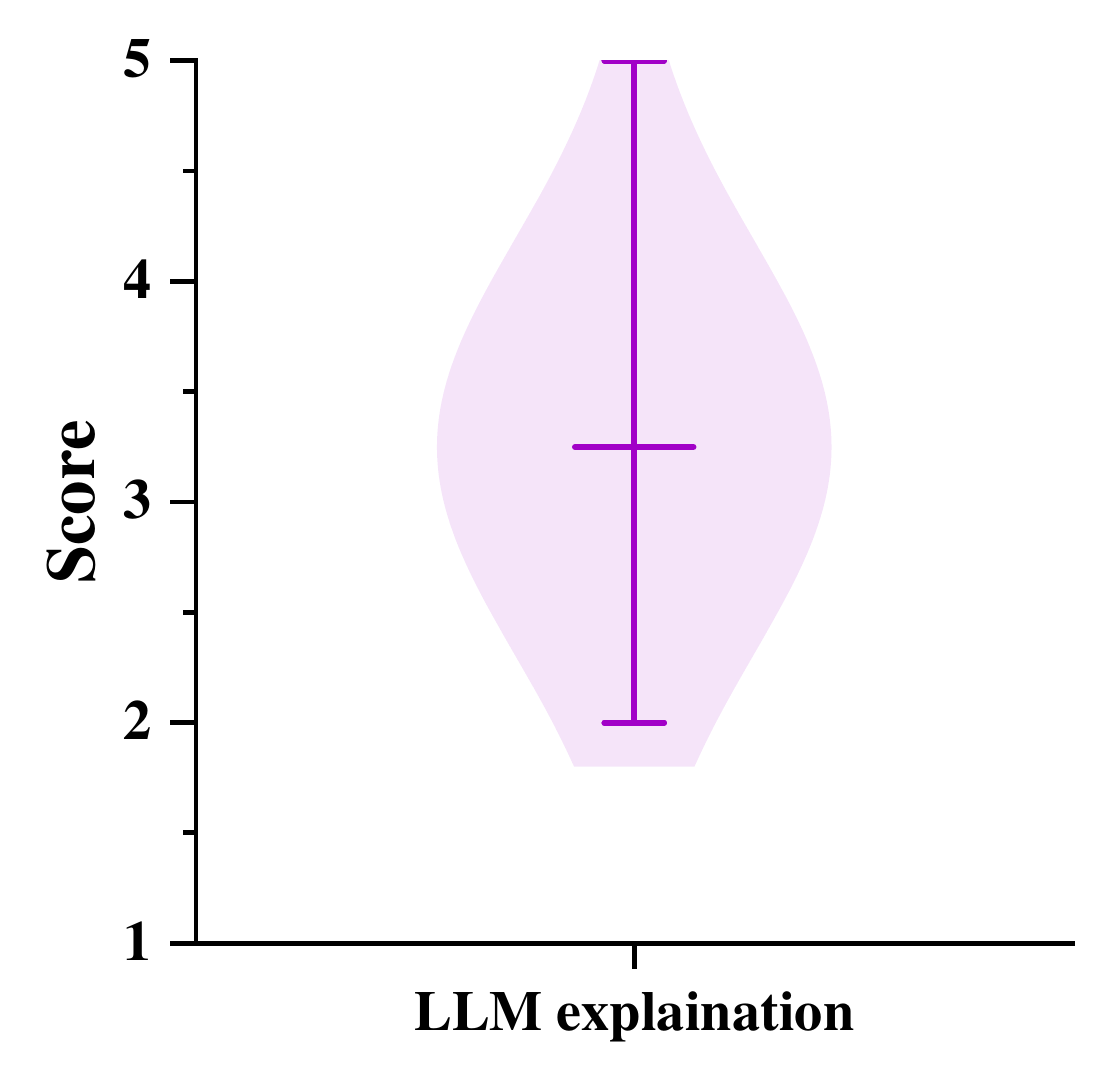}
    \caption{Quality assessment of GPT-4-Turbo explanations.}
    \vspace{6mm}

    \label{fig:Dataset Evaluation}
\end{figure}
\vspace{-6mm}

\subsection{Dataset Quality Evaluation}\label{eval:Dataset Quality Evaluation}

We assessed the quality of assembly code explanations generated by GPT-4-Turbo from both pretraining and finetuning datasets, focusing on their accuracy and relevance to the assembly code.

10 binary domain experts blindly evaluated 50 randomly sampled assembly-explanation pairs using a 5-point scale (1: no explanation, 5: accurate and complete explanation). Results are shown in Fig. \ref{fig:Dataset Evaluation}.

\smallskip
\noindent
\textbf{The quality of assembly-language explanations generated by GPT-4-Turbo is robust.} As depicted in Fig. \ref{fig:Dataset Evaluation}, GPT-4-Turbo achieves 3.65/5, indicating that its explanations fall within the acceptable range of expert expectations. This finding aligns with the results from \citep{zhuo2024bigcodebench}, confirming our data engine produces semantically grounded explanations, which underpins the effectiveness of our datasets.

\vspace{-2mm}

\subsection{Human Assessments}\label{eval:Human Assessments}

\begin{figure}
    \centering 
    \includegraphics[width=1\linewidth]{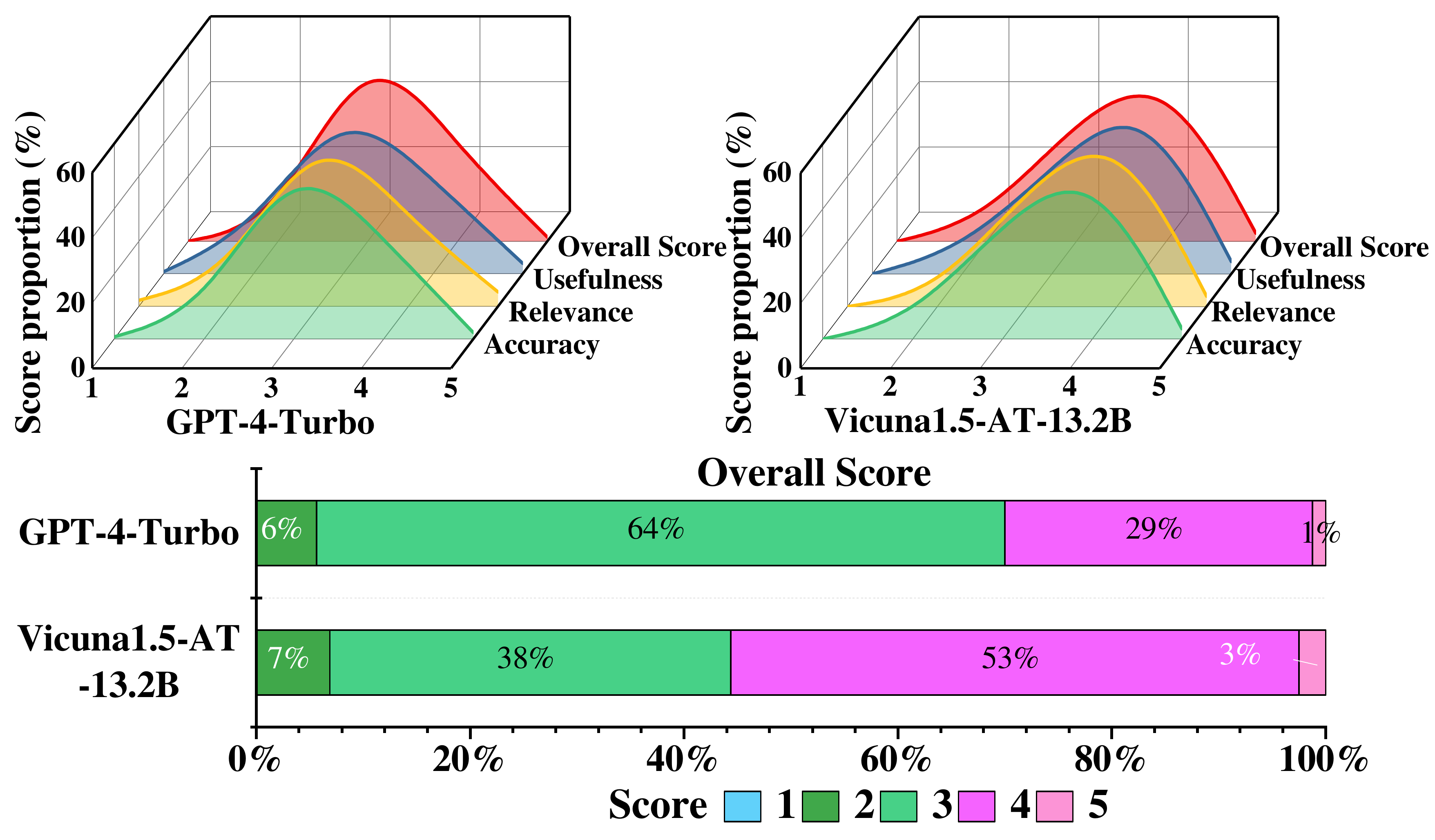}
    \caption{Human assessments results.}
    \label{fig:Human Assess}
    \vspace{4mm}

\end{figure}

We conducted blind assessments with 20 binary and LLM domain experts to evaluate 20 randomly selected assembly code samples. Each sample received two detailed descriptions generated by Vicuna1.5-\textbf{AT}-13.2B and GPT-4-Turbo under identical prompts. Evaluations used a 5-point scale across four metrics: Overall Score, Usefulness, Relevance, and Accuracy. Results are shown in Fig. \ref{fig:Human Assess}.

\smallskip
\noindent
\textbf{Vicuna1.5-\textbf{AT}-13.2B surpasses GPT-4-Turbo in human evaluations.} As shown in Fig. \ref{fig:Human Assess}, Vicuna1.5-\textbf{AT}-13.2B outperformed GPT-4-Turbo by 9\% in the overall score, with consistent advantages across all specific metrics. This human-verified superiority complements the automated metrics in Sec. \ref{Main result 2}, demonstrating \toolname's effectiveness in generating assembly analyses that align with expert judgment.


%% file: tables/Ablation.tex
\begin{table*}[t]
\begin{minipage}[t]{0.5\linewidth}
    \centering
    \caption{Ablation study on ASMA-Bench, w/o means without.}

    \scriptsize
    \setlength{\tabcolsep}{2pt}
    \begin{tabular}{@{}l|cccc@{}}
        \toprule
        \multicolumn{1}{l|}{\textbf{Training data}} &
        \multicolumn{1}{c}{Conversation} &
        \multicolumn{1}{c}{Detail description} &
        \multicolumn{1}{c}{Complex reasoning} &
        \multicolumn{1}{c}{All} \\ \midrule
        Full data   &  90.10  &85.15  & 86.12 & 87.12  \\
        w/o Conversation       &   21.60  & 58.52       &     42.43    &  40.81 \\
        w/o Detailed description &   43.73   &          27.08 &       52.19   &  41.00  \\
        w/o Complex Reasoning  &   57.23   &        53.81     &     21.40  &  44.15  \\
        w/o Instruction Tuning & 12.30  & 15.60    &    11.00   & 12.97 \\ \bottomrule
    \end{tabular}
    \label{tab:ablation1}
\end{minipage}
\begin{minipage}[t]{0.5\linewidth}
    \centering
    \caption{Ablation study on BCSD task.}

    \scriptsize
    \setlength{\tabcolsep}{5pt}
    \begin{tabular}{@{}l|l@{}}
        \toprule
        \multicolumn{1}{l|}{\textbf{Configuration}} &
        \multicolumn{1}{c}{Recall@1(\%) } \\ \midrule
        Full training     &  50.86   \\
        w/o Assembly encoder &   24.41 (↓26.45) \\
        w/o Pre-training &12.02 (↓38.84)\\
        w/o Fine-tuning     &  7.57 \space(↓43.29) \\
        7B Model size     &  45.82 (↓5.04) \\ \bottomrule
    \end{tabular}
    \label{tab:ablation2}
\end{minipage}
\end{table*}

%% file: sec/6_conclusion.tex
\vspace{-2mm}

\section{Conclusion}
We present \toolname, a framework for joint assembly language comprehension and instruction-following. The method improves general-purpose LLMs' assembly code understanding and instruction following while retaining their native code generation capabilities. Experimental results demonstrate the \toolname enhanced models' superior performance across multiple benchmarks. Future work includes enhancing assembly semantic understanding and expanding architectural adaptability.

%% file: main.bbl
\begin{thebibliography}{44}
\providecommand{\natexlab}[1]{#1}
\providecommand{\url}[1]{\texttt{#1}}
\expandafter\ifx\csname urlstyle\endcsname\relax
  \providecommand{\doi}[1]{doi: #1}\else
  \providecommand{\doi}{doi: \begingroup \urlstyle{rm}\Url}\fi

\bibitem[Int(2023)]{Introducing-2023-04-25}
Introducing ai-powered insights in threat intelligence | google cloud blog.
\newblock \url{https://cloud.google.com/blog/products/identity-security/rsa-introducing-ai-powered-insights-threat-intelligence}, 2023.
\newblock Accessed: 2023-04-25.

\bibitem[VT-(2023)]{VT-2023-05-12}
Vt code insight: Updates and q\&a on purpose, challenges, and evolution ~ virustotal blog.
\newblock \url{https://blog.virustotal.com/2023/05/vt-code-insight-updates-and-q-on.html}, 2023.
\newblock Accessed: 2023-05-12.

\bibitem[AUR(2025)]{AUR-(en)-Home}
Aur (en) - home.
\newblock \url{https://aur.archlinux.org/}, 2025.
\newblock Accessed: 2025-05-06.

\bibitem[Arc(2025)]{Arch-Linux-Package-Search}
Arch linux - package search.
\newblock \url{https://archlinux.org/packages/}, 2025.
\newblock Accessed: 2025-05-06.

\bibitem[Mic(2025)]{Microsoft-2025-05-06}
Microsoft security copilot | microsoft security.
\newblock \url{https://www.microsoft.com/en-us/security/business/ai-machine-learning/microsoft-security-copilot}, 2025.
\newblock Accessed: 2025-05-06.

\bibitem[Achiam et~al.(2023)Achiam, Adler, Agarwal, Ahmad, Akkaya, Aleman, Almeida, Altenschmidt, Altman, Anadkat, et~al.]{achiam2023gpt}
J.~Achiam, S.~Adler, S.~Agarwal, L.~Ahmad, I.~Akkaya, F.~L. Aleman, D.~Almeida, J.~Altenschmidt, S.~Altman, S.~Anadkat, et~al.
\newblock Gpt-4 technical report.
\newblock \emph{arXiv preprint arXiv:2303.08774}, 2023.

\bibitem[Al-Kaswan et~al.(2023)Al-Kaswan, Ahmed, Izadi, Sawant, Devanbu, and van Deursen]{al2023extending}
A.~Al-Kaswan, T.~Ahmed, M.~Izadi, A.~A. Sawant, P.~Devanbu, and A.~van Deursen.
\newblock Extending source code pre-trained language models to summarise decompiled binaries.
\newblock In \emph{2023 IEEE International Conference on Software Analysis, Evolution and Reengineering (SANER)}, pages 260--271. IEEE, 2023.

\bibitem[Boland and Black(2012)]{boland2012juliet}
T.~Boland and P.~E. Black.
\newblock Juliet 1. 1 c/c++ and java test suite.
\newblock \emph{Computer}, 45\penalty0 (10):\penalty0 88--90, 2012.

\bibitem[Brooks et~al.(2023)Brooks, Holynski, and Efros]{brooks2023instructpix2pix}
T.~Brooks, A.~Holynski, and A.~A. Efros.
\newblock Instructpix2pix: Learning to follow image editing instructions.
\newblock In \emph{Proceedings of the IEEE/CVF Conference on Computer Vision and Pattern Recognition}, pages 18392--18402, 2023.

\bibitem[Chen et~al.(2021)Chen, Tworek, Jun, Yuan, Pinto, Kaplan, Edwards, Burda, Joseph, Brockman, et~al.]{chen2021evaluating}
M.~Chen, J.~Tworek, H.~Jun, Q.~Yuan, H.~P. D.~O. Pinto, J.~Kaplan, H.~Edwards, Y.~Burda, N.~Joseph, G.~Brockman, et~al.
\newblock Evaluating large language models trained on code.
\newblock \emph{arXiv preprint arXiv:2107.03374}, 2021.

\bibitem[Chiang et~al.(2023)Chiang, Li, Lin, Sheng, Wu, Zhang, Zheng, Zhuang, Zhuang, Gonzalez, et~al.]{chiang2023vicuna}
W.-L. Chiang, Z.~Li, Z.~Lin, Y.~Sheng, Z.~Wu, H.~Zhang, L.~Zheng, S.~Zhuang, Y.~Zhuang, J.~E. Gonzalez, et~al.
\newblock Vicuna: An open-source chatbot impressing gpt-4 with 90\%* chatgpt quality.
\newblock \emph{See https://vicuna. lmsys. org (accessed 14 April 2023)}, 2\penalty0 (3):\penalty0 6, 2023.

\bibitem[Chlipala(2011)]{chlipala2011mostly}
A.~Chlipala.
\newblock Mostly-automated verification of low-level programs in computational separation logic.
\newblock In \emph{Proceedings of the 32nd ACM SIGPLAN conference on Programming language design and implementation}, pages 234--245, 2011.

\bibitem[Chowdhery et~al.(2023)Chowdhery, Narang, Devlin, Bosma, Mishra, Roberts, Barham, Chung, Sutton, Gehrmann, et~al.]{chowdhery2023palm}
A.~Chowdhery, S.~Narang, J.~Devlin, M.~Bosma, G.~Mishra, A.~Roberts, P.~Barham, H.~W. Chung, C.~Sutton, S.~Gehrmann, et~al.
\newblock Palm: Scaling language modeling with pathways.
\newblock \emph{Journal of Machine Learning Research}, 24\penalty0 (240):\penalty0 1--113, 2023.

\bibitem[Chung et~al.(2024)Chung, Hou, Longpre, Zoph, Tay, Fedus, Li, Wang, Dehghani, Brahma, et~al.]{chung2024scaling}
H.~W. Chung, L.~Hou, S.~Longpre, B.~Zoph, Y.~Tay, W.~Fedus, Y.~Li, X.~Wang, M.~Dehghani, S.~Brahma, et~al.
\newblock Scaling instruction-finetuned language models.
\newblock \emph{Journal of Machine Learning Research}, 25\penalty0 (70):\penalty0 1--53, 2024.

\bibitem[Cummins et~al.(2024)Cummins, Seeker, Grubisic, Roziere, Gehring, Synnaeve, and Leather]{cummins2024meta}
C.~Cummins, V.~Seeker, D.~Grubisic, B.~Roziere, J.~Gehring, G.~Synnaeve, and H.~Leather.
\newblock Meta large language model compiler: Foundation models of compiler optimization.
\newblock \emph{arXiv preprint arXiv:2407.02524}, 2024.

\bibitem[Ding et~al.(2019)Ding, Fung, and Charland]{ding2019asm2vec}
S.~H. Ding, B.~C. Fung, and P.~Charland.
\newblock Asm2vec: Boosting static representation robustness for binary clone search against code obfuscation and compiler optimization.
\newblock In \emph{2019 ieee symposium on security and privacy (sp)}, pages 472--489. IEEE, 2019.

\bibitem[Floridi and Chiriatti(2020)]{floridi2020gpt}
L.~Floridi and M.~Chiriatti.
\newblock Gpt-3: Its nature, scope, limits, and consequences.
\newblock \emph{Minds and Machines}, 30:\penalty0 681--694, 2020.

\bibitem[Guo et~al.(2024)Guo, Zhu, Yang, Xie, Dong, Zhang, Chen, Bi, Wu, Li, et~al.]{guo2024deepseek}
D.~Guo, Q.~Zhu, D.~Yang, Z.~Xie, K.~Dong, W.~Zhang, G.~Chen, X.~Bi, Y.~Wu, Y.~Li, et~al.
\newblock Deepseek-coder: When the large language model meets programming--the rise of code intelligence.
\newblock \emph{arXiv preprint arXiv:2401.14196}, 2024.

\bibitem[Hui et~al.(2024)Hui, Yang, Cui, Yang, Liu, Zhang, Liu, Zhang, Yu, Lu, et~al.]{hui2024qwen2}
B.~Hui, J.~Yang, Z.~Cui, J.~Yang, D.~Liu, L.~Zhang, T.~Liu, J.~Zhang, B.~Yu, K.~Lu, et~al.
\newblock Qwen2. 5-coder technical report.
\newblock \emph{arXiv preprint arXiv:2409.12186}, 2024.

\bibitem[Izacard et~al.(2021)Izacard, Caron, Hosseini, Riedel, Bojanowski, Joulin, and Grave]{izacard2021unsupervised}
G.~Izacard, M.~Caron, L.~Hosseini, S.~Riedel, P.~Bojanowski, A.~Joulin, and E.~Grave.
\newblock Unsupervised dense information retrieval with contrastive learning.
\newblock \emph{arXiv preprint arXiv:2112.09118}, 2021.

\bibitem[Jiang et~al.(2023)Jiang, Wang, Liu, Xu, Tan, and Zhang]{jiang2023nova}
N.~Jiang, C.~Wang, K.~Liu, X.~Xu, L.~Tan, and X.~Zhang.
\newblock Nova: Generative language models for binaries.
\newblock \emph{arXiv preprint arXiv:2311.13721}, 2023.

\bibitem[Jin et~al.(2023)Jin, Larson, Yang, and Lin]{jin2023binary}
X.~Jin, J.~Larson, W.~Yang, and Z.~Lin.
\newblock Binary code summarization: Benchmarking chatgpt/gpt-4 and other large language models.
\newblock \emph{arXiv preprint arXiv:2312.09601}, 2023.

\bibitem[Karg{\'e}n and Shahmehri(2017)]{kargen2017towards}
U.~Karg{\'e}n and N.~Shahmehri.
\newblock Towards robust instruction-level trace alignment of binary code.
\newblock In \emph{2017 32nd IEEE/ACM International Conference on Automated Software Engineering (ASE)}, pages 342--352. IEEE, 2017.

\bibitem[Kim et~al.(2023)Kim, Moon, Tabrizi, Lee, Mahoney, Keutzer, and Gholami]{kim2023llm}
S.~Kim, S.~Moon, R.~Tabrizi, N.~Lee, M.~W. Mahoney, K.~Keutzer, and A.~Gholami.
\newblock An llm compiler for parallel function calling.
\newblock \emph{arXiv preprint arXiv:2312.04511}, 2023.

\bibitem[Liu et~al.(2024)Liu, Li, Wu, and Lee]{liu2024visual}
H.~Liu, C.~Li, Q.~Wu, and Y.~J. Lee.
\newblock Visual instruction tuning.
\newblock \emph{Advances in neural information processing systems}, 36, 2024.

\bibitem[Mantovani et~al.(2022)Mantovani, Compagna, Shoshitaishvili, and Balzarotti]{mantovani2022convergence}
A.~Mantovani, L.~Compagna, Y.~Shoshitaishvili, and D.~Balzarotti.
\newblock The convergence of source code and binary vulnerability discovery--a case study.
\newblock In \emph{Proceedings of the 2022 ACM on Asia Conference on Computer and Communications Security}, pages 602--615, 2022.

\bibitem[Megira et~al.(2018)Megira, Pangesti, and Wibowo]{megira2018malware}
S.~Megira, A.~Pangesti, and F.~Wibowo.
\newblock Malware analysis and detection using reverse engineering technique.
\newblock In \emph{Journal of Physics: Conference Series}, volume 1140, page 012042. IOP Publishing, 2018.

\bibitem[Ouyang et~al.(2022)Ouyang, Wu, Jiang, Almeida, Wainwright, Mishkin, Zhang, Agarwal, Slama, Ray, et~al.]{ouyang2022training}
L.~Ouyang, J.~Wu, X.~Jiang, D.~Almeida, C.~Wainwright, P.~Mishkin, C.~Zhang, S.~Agarwal, K.~Slama, A.~Ray, et~al.
\newblock Training language models to follow instructions with human feedback.
\newblock \emph{Advances in neural information processing systems}, 35:\penalty0 27730--27744, 2022.

\bibitem[Panagopoulou et~al.(2023)Panagopoulou, Xue, Yu, Li, Li, Joty, Xu, Savarese, Xiong, and Niebles]{panagopoulou2023x}
A.~Panagopoulou, L.~Xue, N.~Yu, J.~Li, D.~Li, S.~Joty, R.~Xu, S.~Savarese, C.~Xiong, and J.~C. Niebles.
\newblock X-instructblip: A framework for aligning x-modal instruction-aware representations to llms and emergent cross-modal reasoning.
\newblock \emph{arXiv preprint arXiv:2311.18799}, 2023.

\bibitem[Pei et~al.(2020)Pei, Xuan, Yang, Jana, and Ray]{pei2020trex}
K.~Pei, Z.~Xuan, J.~Yang, S.~Jana, and B.~Ray.
\newblock Trex: Learning execution semantics from micro-traces for binary similarity.
\newblock \emph{arXiv preprint arXiv:2012.08680}, 2020.

\bibitem[Ram et~al.(2023)Ram, Levine, Dalmedigos, Muhlgay, Shashua, Leyton-Brown, and Shoham]{ram2023context}
O.~Ram, Y.~Levine, I.~Dalmedigos, D.~Muhlgay, A.~Shashua, K.~Leyton-Brown, and Y.~Shoham.
\newblock In-context retrieval-augmented language models.
\newblock \emph{Transactions of the Association for Computational Linguistics}, 11:\penalty0 1316--1331, 2023.

\bibitem[Roziere et~al.(2023)Roziere, Gehring, Gloeckle, Sootla, Gat, Tan, Adi, Liu, Sauvestre, Remez, et~al.]{roziere2023code}
B.~Roziere, J.~Gehring, F.~Gloeckle, S.~Sootla, I.~Gat, X.~E. Tan, Y.~Adi, J.~Liu, R.~Sauvestre, T.~Remez, et~al.
\newblock Code llama: Open foundation models for code.
\newblock \emph{arXiv preprint arXiv:2308.12950}, 2023.

\bibitem[Shang et~al.(2024)Shang, Cheng, Chen, Zhang, Hu, Yu, Li, Zhang, and Yu]{shang2024far}
X.~Shang, S.~Cheng, G.~Chen, Y.~Zhang, L.~Hu, X.~Yu, G.~Li, W.~Zhang, and N.~Yu.
\newblock How far have we gone in binary code understanding using large language models.
\newblock In \emph{2024 IEEE International Conference on Software Maintenance and Evolution (ICSME)}, pages 1--12. IEEE, 2024.

\bibitem[Su et~al.(2024)Su, Xu, Huang, Zhang, Ye, Huang, and Zhang]{su2024codeart}
Z.~Su, X.~Xu, Z.~Huang, Z.~Zhang, Y.~Ye, J.~Huang, and X.~Zhang.
\newblock Codeart: Better code models by attention regularization when symbols are lacking.
\newblock \emph{Proceedings of the ACM on Software Engineering}, 1\penalty0 (FSE):\penalty0 562--585, 2024.

\bibitem[Tan et~al.(2024)Tan, Luo, Li, and Zhang]{tan2024llm4decompile}
H.~Tan, Q.~Luo, J.~Li, and Y.~Zhang.
\newblock Llm4decompile: Decompiling binary code with large language models.
\newblock \emph{arXiv preprint arXiv:2403.05286}, 2024.

\bibitem[Taviss et~al.(2024)Taviss, Ding, Zulkernine, Charland, and Acharya]{taviss2024asm2seq}
S.~Taviss, S.~H. Ding, M.~Zulkernine, P.~Charland, and S.~Acharya.
\newblock Asm2seq: Explainable assembly code functional summary generation for reverse engineering and vulnerability analysis.
\newblock \emph{Digital Threats: Research and Practice}, 5\penalty0 (1):\penalty0 1--25, 2024.

\bibitem[Thirumoorthy et~al.(2022)]{thirumoorthy2022feature}
K.~Thirumoorthy et~al.
\newblock A feature selection model for software defect prediction using binary rao optimization algorithm.
\newblock \emph{Applied Soft Computing}, 131:\penalty0 109737, 2022.

\bibitem[Touvron et~al.(2023)Touvron, Lavril, Izacard, Martinet, Lachaux, Lacroix, Rozi{\`e}re, Goyal, Hambro, Azhar, et~al.]{touvron2023llama}
H.~Touvron, T.~Lavril, G.~Izacard, X.~Martinet, M.-A. Lachaux, T.~Lacroix, B.~Rozi{\`e}re, N.~Goyal, E.~Hambro, F.~Azhar, et~al.
\newblock Llama: Open and efficient foundation language models.
\newblock \emph{arXiv preprint arXiv:2302.13971}, 2023.

\bibitem[Wang et~al.(2022)Wang, Qu, Katz, Zhu, Gao, Qiu, Zhuge, and Zhang]{wang2022jtrans}
H.~Wang, W.~Qu, G.~Katz, W.~Zhu, Z.~Gao, H.~Qiu, J.~Zhuge, and C.~Zhang.
\newblock Jtrans: Jump-aware transformer for binary code similarity detection.
\newblock In \emph{Proceedings of the 31st ACM SIGSOFT International Symposium on Software Testing and Analysis}, pages 1--13, 2022.

\bibitem[Wang et~al.(2024)Wang, Gao, Zhang, Sha, Sun, Zhou, Zhu, Sun, Qiu, and Xiao]{wang2024clap}
H.~Wang, Z.~Gao, C.~Zhang, Z.~Sha, M.~Sun, Y.~Zhou, W.~Zhu, W.~Sun, H.~Qiu, and X.~Xiao.
\newblock Clap: Learning transferable binary code representations with natural language supervision.
\newblock In \emph{Proceedings of the 33rd ACM SIGSOFT International Symposium on Software Testing and Analysis}, pages 503--515, 2024.

\bibitem[Wang et~al.(2021)Wang, Wang, Joty, and Hoi]{wang2021codet5}
Y.~Wang, W.~Wang, S.~Joty, and S.~C. Hoi.
\newblock Codet5: Identifier-aware unified pre-trained encoder-decoder models for code understanding and generation.
\newblock \emph{arXiv preprint arXiv:2109.00859}, 2021.

\bibitem[Xiong et~al.(2023)Xiong, Chen, Chen, Gao, Cheng, and Zhang]{xiong2023hext5}
J.~Xiong, G.~Chen, K.~Chen, H.~Gao, S.~Cheng, and W.~Zhang.
\newblock Hext5: Unified pre-training for stripped binary code information inference.
\newblock In \emph{2023 38th IEEE/ACM International Conference on Automated Software Engineering (ASE)}, pages 774--786. IEEE, 2023.

\bibitem[Zhu et~al.(2024)Zhu, Guo, Shao, Yang, Wang, Xu, Wu, Li, Gao, Ma, et~al.]{zhu2024deepseek}
Q.~Zhu, D.~Guo, Z.~Shao, D.~Yang, P.~Wang, R.~Xu, Y.~Wu, Y.~Li, H.~Gao, S.~Ma, et~al.
\newblock Deepseek-coder-v2: Breaking the barrier of closed-source models in code intelligence.
\newblock \emph{arXiv preprint arXiv:2406.11931}, 2024.

\bibitem[Zhuo et~al.(2024)Zhuo, Vu, Chim, Hu, Yu, Widyasari, Yusuf, Zhan, He, Paul, et~al.]{zhuo2024bigcodebench}
T.~Y. Zhuo, M.~C. Vu, J.~Chim, H.~Hu, W.~Yu, R.~Widyasari, I.~N.~B. Yusuf, H.~Zhan, J.~He, I.~Paul, et~al.
\newblock Bigcodebench: Benchmarking code generation with diverse function calls and complex instructions.
\newblock \emph{arXiv preprint arXiv:2406.15877}, 2024.

\end{thebibliography}
